\documentclass[twocolumn,showpacs,preprintnumbers,amsmath,amssymb]{revtex4}

\usepackage{graphicx}
\usepackage{dcolumn}
\usepackage{bm}

\begin{document}

\preprint{APS/123-QED}

\title{Local Tunneling Study of Three-Dimensional
Order Parameter in the $\pi$-band of Al-doped MgB$_2$ Single Crystals}

\author{F. Giubileo}
\author{F. Bobba}%
\author{A. Scarfato}
\author{A.M. Cucolo}%
\affiliation{%
CNR-INFM Laboratorio Regionale SUPERMAT e Dipartimento di Fisica
"E.R. Caianiello", Universit$\grave{a}$ degli Studi di Salerno,
via Salvador Allende, 84081 Baronissi (SA), Italy.
}%

\author{A. Kohen}
\author{D. Roditchev}
\affiliation{Institut des Nanosciences de Paris, INSP,
Universit$\acute{e}$ P. et M. Curie Paris 6, CNRS, UMR 75-88,
Paris, France
}%

\author{N. Zhigadlo}
\author{J. Karpinski}%
\affiliation{Solid State Physics Laboratory, ETH Zurich, CH-8093
Zurich, Switzerland}%


\begin{abstract}
We have performed local tunneling spectroscopy on high quality
Mg$_{1-x}$Al$_x$B$_2$ single crystals by means of Variable
Temperature Scanning Tunneling Spectroscopy (STS) in magnetic
field up to 3 Tesla. Single gap conductance spectra due to
$c$-axis tunneling were extensively measured, probing different
amplitudes of the three-dimensional $\Delta_\pi$ as a function of
Al content. Temperature and magnetic field dependences of the
conductance spectra were studied in S-I-N configuration: the
effect of the doping resulted in a monotonous reduction of the
locally measured $T_C$ down to 24K for x=0.2. On the other hand,
we have found that the gap amplitude shows a maximum value
$\Delta_\pi= 2.3$ meV for x=0.1, while the $\Delta_\pi / T_C$
ratio increases monotonously with doping. The locally measured
upper critical field was found to be strongly related to the gap
amplitude, showing the maximum value $H_{c2}\simeq3T$ for x=0.1
substituted samples. For this Al concentration the data revealed
some spatial inhomogeneity in the distribution of $\Delta_\pi$ on
nanometer scale.
\end{abstract}

\maketitle

Five years after Nagamatsu et al. \cite{Nagamatsu} reported
MgB$_2$ to be superconductor, the huge worldwide experimental and
theoretical effort seems to have established the main features of
superconductivity in this compound. Indeed, the strong electronic
coupling to the high-frequency in-plane boron modes ($E_{2g}$  at
the zone centre $\Gamma$) and the number of holes at the Fermi
level in the $\sigma$ bands are able to explain  a transition
temperature $T_C$ as high as 39 K \cite{Chol,An}. Moreover, it is
now demonstrated
\cite{Kortus,Budko,Liu,Giubileo1,GiubileoPRL,Szabo,Gonnelli1,Iavarone1,PhysC}
that MgB$_2$ is a two-gap superconductor with two distinct energy
gaps: a large gap $\Delta_\sigma$ originating from two-dimensional
(2D) $\sigma$ bands and a small gap $\Delta_\pi$ originating from
three-dimensional (3D) $\pi$ bands. The presence of two bands with
distinct superconducting gaps leads to several unusual properties,
like the temperature and field dependent anisotropy which dominate
the magnetic and transport properties. Anisotropy is related to
the intraband and interband electron scattering that can be
modified by partial chemical substitutions. In particular,
aluminium (replacing magnesium) \cite{sluski}, and carbon
(replacing boron) \cite{ribeiro} have successfully entered in the
MgB$_2$ structure, doping the material with additional electrons:
small variations of the interband scattering have been predicted
for C substitutions, while it has been demonstrated that Al doping
can realize a considerable out-of-plane distortions of the B atoms
\cite{Erwin} causing a significant increase of the interband
scattering with consequent increasing of $\Delta_\pi$ and
decreasing of $\Delta_\sigma$ \cite{BH}.

Experimentally, it has been observed that the superconducting
transition temperature of both Mg$_{1-x}$Al$_x$B$_2$ and
Mg(B$_{1-y}$C$_y$)$_2$ decreases with doping \cite{sluski,
bianconi} and in the case of Al (C), superconductivity disappears
for $x>0.5$ ($y>0.3$) \cite{postorino,renker}. Recently,
measurements of the amplitude of the energy gaps  have been
performed by means of different techniques (specific heat, point
contact, STM) on Al doped \cite{PuttiAff,PuttiGon} as well as on
neutron irradiated polycrystals \cite{Wang,Putti06} and on
disordered thin films \cite{Iavarone}. From these studies a quite
general trend seems to relate the variation of both energy gaps
with $T_C$, however a different behavior of $\Delta_\pi$ has been
reported for Al-doped single-crystals, indicating large gap values
for doping levels up to 10\% and quite small values for higher
doping levels \cite{Gonnelli,Karpinski}. Results on C-doped
samples also are controversial and the analysis of the whole set
of data resulted in an extended debate
\cite{Kortus2,reply1,reply2} still waiting for a definite answer.
It is our opinion that in some cases, disagreement arises due to
the non-local nature of used experimental techniques and to the
high number of fitting parameters necessary to reproduce the
experiments.



In this paper we report a systematic study performed by Scanning
Tunneling Spectroscopy (STS)  on high quality
Mg$_{1-x}$Al$_{x}$B$_2$ single crystals, for different Al
concentrations. Directional tunneling along $c$-axis allowed us to
selectively probe the $\pi$ band energy gap, with high spatial and
energy resolution. In particular, by measuring the temperature
dependence of the tunneling spectra, the local $T_C$ was inferred,
corresponding to the  energy gap measured in the same location.
The magnetic field dependence was also studied to evidence
correlations of the locally measured upper critical field $H_{c2}$
with the gap amplitude. Moreover, the high spatial resolution of
the STS technique allowed to evidence possible non-homogeneities
of the superconducting properties on the sample surface with
variation of $\Delta_\pi$ depending on the doping.

Single crystals of Mg$_{1-x}$Al$_x$B$_2$ were grown by high
pressure method in a cubic-anvil press in the same way as the pure
crystals \cite{13}. The STS experiments were carried out on
crystals with nominal 0$\%$, 10$\%$ and 20$\%$ Al content by means
of an UHV variable temperature STM. The tunneling junctions were
achieved by approaching a mechanically etched Pt/Ir tip to the
$c$-axis oriented surface of the crystals. As expected, the STS
measurements revealed only a single gap structure in the dI/dV
spectra at low temperatures because the probability for direct
tunneling into the 3D-sheet of the Fermi surface results much
higher than the probability for tunneling into the 2D-part of the
Fermi surface which has no states with wavevector parallel to the
$c$-axis.

\begin{figure}
\includegraphics{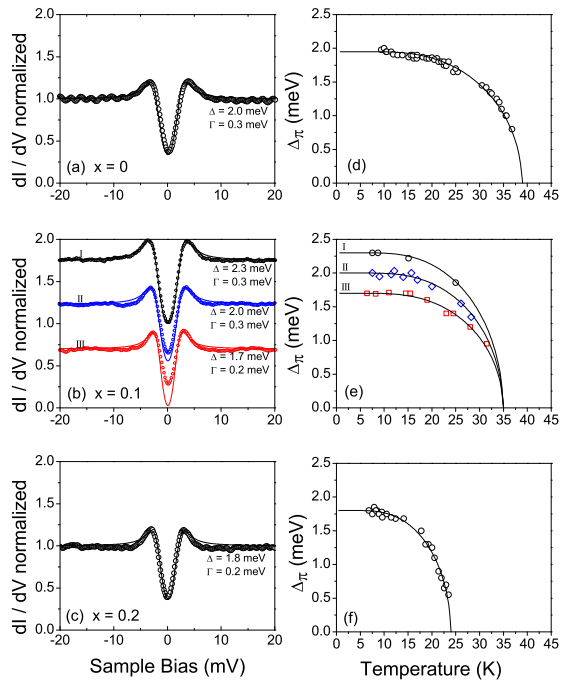}
\caption{\label{fig:epsart} Left plots: Low temperature spectra
measured in Mg$_{1-x}$Al$_x$B$_2$ for x = 0 (a), x=0.1 (b) and
x=0.2 (c). Solid lines represent the theoretical fittings
calculated by considering a single gap isotropic BCS density of
states with a smearing $\Gamma$ parameter. Right plots:
corresponding temperature dependence of the superconducting energy
gap $\Delta_\pi$ as extracted from the theoretical fittings.
Experimental data are compared to the theoretical BCS behavior
(solid lines). }
\end{figure}


In Fig. 1a,b,c we show the dI/dV characteristics measured at $T =
6.5 K$ respectively on pure MgB$_2$, and on samples with 10\%
(x=0.1) and 20\% (x=0.2) Al content. It can be observed that all
the tunneling spectra are well reproduced by an isotropic BCS
state density with a single gap value $\Delta_{\pi}$,
corresponding to the 3D $\pi$-band, and a phenomenological
smearing factor $\Gamma$, corresponding to finite lifetime of the
quasiparticles, as introduced by Dynes \cite{Dynes}. These are the
only two fitting parameters needed to model the experimental data
while the temperature was directly measured. The experiments
indicated that pure MgB$_2$ crystals were highly homogeneous with
the sample surface characterized by a superconducting energy gap
$\Delta_\pi = 2.00 \pm 0.05$ meV, i.e. with less than 3\% spread
in the  values measured in different locations.

The behavior of the doped crystals appeared to be quite different.
For the x=0.1 substituted crystals, the 3D $\Delta_\pi$ resulted
to be non-homogeneous in its spatial distribution on nanometer
scale, with values varying between 1.5 meV $< \Delta_\pi <$ 2.3
meV, as observed in Fig. 1b referring to different locations of
the same sample. The spectrum signed (I) for which we found
$\Delta_\pi$ = 2.3 meV, was the statistically most present in
about 90\% of the locations. However, in few cases, we have
measured different gap amplitudes as observed in curve (II) with
$\Delta_\pi$ = 2.0 meV and in curve (III) with $\Delta_\pi$ = 1.7
meV. The energy gap variations in the x=0.1 substituted samples,
can be due to different local Al concentrations arising during the
crystal growth process. Indeed, structural changes can occur in
crystals when the Al content is increased beyond a critical value
$x \simeq 0.1$. These changes include the segregation of a
non-superconducting, Al-rich phase and the formation of
superstructures along the $c$-axis \cite{Karpinski}. We notice
that the most satisfactory agreement between theory and
experiments was obtained for the spectra statistically more
present characterized by the largest value of the energy gap,
$\Delta_\pi = 2.3$meV, corresponding to a 15\% increase of the
superconducting energy gap compared to the case of pure MgB$_2$.
\begin{figure*}
\includegraphics[width= 15.8 cm]{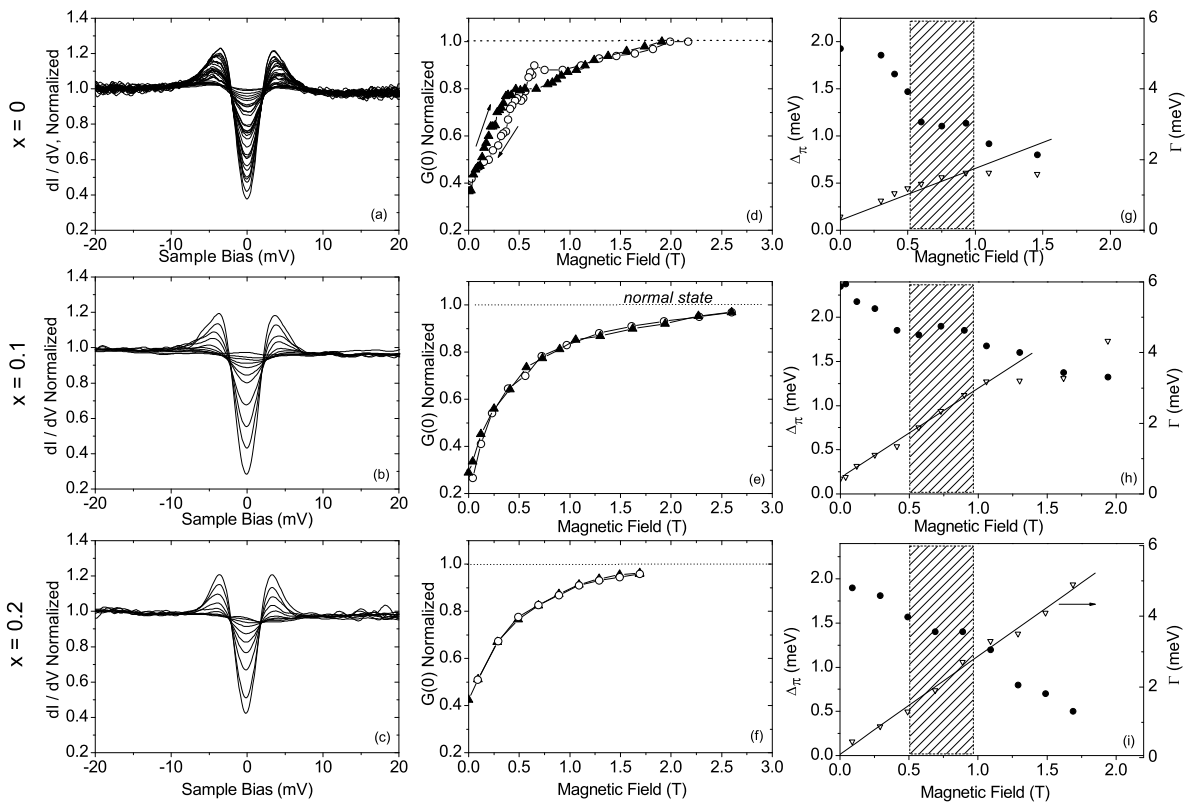}
\caption{\label{fig:wide}Magnetic field dependence of the
conductance spectra measured at $T\simeq6.5K$. Different doping
levels are reported on different rows. In the first column the
tunneling conductance spectra measured for rising fields are
reported. In the central column the field evolution of the ZBC,
for increasing (solid triangles) and decreasing (empty circles)
fields, are shown. In the third column the magnetic field
dependence of the 3D $\Delta_\pi$ as obtained by theoretical
fitting is presented. Solid symbols refer to the energy gap
amplitude, while open symbols refer to the $\Gamma$ values.}
\end{figure*}
For the x=0.2 substituted crystals, statistic in several locations
showed quite homogeneous superconducting properties on the sample
surface. The measured spectra evidenced single gap features (Fig.
1c) with $\Delta_\pi$= 1.8 meV, with less than 6\% spread in the
measured values. We notice that this last estimation results much
higher than what previously reported for similar doping levels
\cite{Gonnelli}.


For all samples, we have performed complete measurements of the
temperature dependence of the tunneling spectra in the range
between 5K and 40K. In Fig. 1d,e,f, the gap amplitude as inferred
from the theoretical fittings is plotted as a function of the
temperature. In the case of pure MgB$_2$ (Fig. 1d), a BCS
dependence (solid line) of the data (scattered symbols) is found
indicating a local $T_C$ = 39K. In the case of 10\% Al-doping
(Fig. 1e), gaps of different amplitudes all vanish at the same
critical temperature T$_C \simeq$ 35 K indicating that variations
of the 3D order parameter in the $\pi$-band occur on a scale less
than the superconducting coherence length. For samples with higher
doping level, x=0.2 (Fig. 1f), a local $T_C$=24K is found.

We also performed a complete analysis of the local response to
external magnetic fields up to 3 T, with the tunneling current and
the applied field parallel to the c-axis of the crystal. The
samples were cooled in zero magnetic field. At low temperature,
the field was slowly increased from zero up to 3 T and then
reduced to zero again, to evidence any hysteretic behavior. Since
the reported spectra were averaged over many vortices passing
under the tip \cite{APL}, the main effect of the magnetic field
was, as expected, the progressive filling of states inside the
energy gap.

In Fig. 2 we show a complete set of data recorded in magnetic
field for x=0 (Fig. 3a,d,g), x=0.1 (Fig. 2b,e,h), and x=0.2 (Fig.
2c,f,i). In the first column we show the evolution of the
normalized tunneling conductance spectra as measured at $T \simeq
6.5 $K. The field dynamics of the DOS at the Fermi level is
reported in the second column where the evolution of the Zero-Bias
Conductance (ZBC) is presented. We notice that for pure crystals
(Fig. 2d), the ZBC rapidly rises for low fields and reaches a
value of about 80\% of the normal state ZBC around 0.4 T. As the
field further increases, the filling of states becomes much
slower, the two different dynamics being separated by an almost
flat crossover region. Finally, the gap fills completely around
2.2 T. By lowering the field we observed a similar behavior, with
the crossover region slightly shifted to higher fields. We
speculate that the crossover region can be associated to the
rotation of the vortex lattice in the pure MgB$_2$ \cite{Qubit},
while the hysteretic behavior seems to indicate different vortex
dynamics for increasing and decreasing fields, which may be due to
geometrical barriers, vortex pinning, and/or lattice
re-arrangements.

In the case of 10\% Al doping (Fig. 2e), the data refer to
locations with $\Delta_\pi = 2.3$ meV. The field dynamics of the
DOS at the Fermi level again shows a rapid rising of the ZBC for
low magnetic fields. However, for increasing fields, the filling
of states tends to saturate and, at 2.5T, it is still possible to
distinguish the presence of the superconducting energy gap in the
measured spectra. Extrapolation of the data in this region leads
to $H_{c2} \simeq$ 3 T, corresponding to a value 30\% higher than
that observed in the case of pure MgB$_2$. For x=0.2, the ZBC
evolution in magnetic field indicates a reduced $H_{c2}\simeq 1.8$
T. We notice that, for both substitutions, the ZBC doesn't show
any hysteretic behavior. Finally, in Fig. 2c,f,i we show  the
magnetic field dependence of $\Delta_\pi$  for the three samples
as inferred from the theoretical fittings. We observe a clear
reduction of the gap amplitude for fields up to 0.5 T followed by
a region between 0.5 T and 1.0 T, in which no significant
variations occur, while pair-breaking continuously increases due
to the applied magnetic field.
This observation seems to suggest that around 0.5T the
contribution to the superconductivity due to the phonon mediated
electron-electron interactions in the $\pi$-band itself is not
efficient anymore, while for higher fields the energy gap survives
due to both the phonon exchange with $\sigma$-band \cite{Suhl}
and/or to the quasiparticle interband scattering.

\begin{table}
\caption{\label{tab:table2}Summary of our STM results.}
\begin{ruledtabular}
\begin{tabular}{ccccc}
 x & $T_C$(K)& $\Delta_\pi (meV)$ & $H_{c2}$(T) &
 2$\Delta_\pi / K_BT_C$\\
\hline 0   & 39 & 2.0 & 2.2 & 1.17  \\
       0.1 & 35 & 2.3 & 3.0 & 1.52  \\
       0.2 & 24 & 1.8 & 1.8 & 1.74  \\
\end{tabular}
\end{ruledtabular}
\end{table}

\begin{figure}[t]
\includegraphics[width= 8.5 cm]{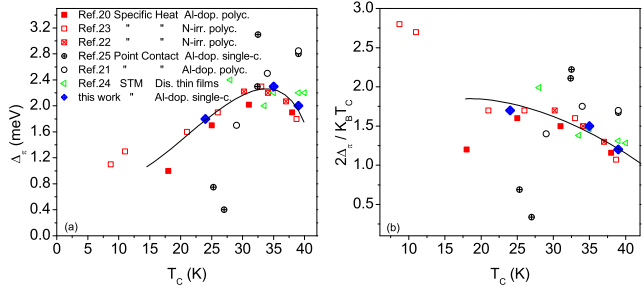}
\caption{\label{fig:epsart3} (a) $\Delta_\pi(0)$ and (b)
$2\Delta_\pi(0)/K_B T_C$ as function of $T_C$, compared with data
 in literature. Same symbols have been used in both plots
and the full lines are guides to the eye. }
\end{figure}
Finally, in Table I we summarize our results that in Fig. 3 are
compared with the literature. The significant spread of the data
reported by different groups is at the origin of the recent, hot
debate \cite{Kortus2,reply1,reply2}, nevertheless a quite general
trend (full lines) for both the 3D energy gap $\Delta_\pi (0)$ and
the $2\Delta_\pi (0) / K_B T_C$ ratio as a function of  $T_C$ can
be inferred regardless to the nature of the measured samples:
doped/irradiated/disordered - single crystals/polycristals/thin
films. The monotonous increase of the $2\Delta_\pi (0) / K_B T_C$
ratio and the maximum value of $\Delta_\pi (0)$ found for x=0.1
appear as a strong confirmation of the hypothesis due to A.
Bussmann-Holder and A. Bianconi \cite{BH} with our data adding a
clear relation between the locally measured values of $\Delta_\pi
(0)$ and $H_{c2}$.

In conclusion, we have performed a systematic study of the local
temperature and magnetic field dependence of the 3D energy gap
$\Delta_\pi$ in Mg$_{1-x}$Al$_{x}$B$_2$ single crystals by means
of Scanning Tunneling Spectroscopy. By working with high quality
single crystals we succeded to selectively measure the behavior of
the only $\Delta_\pi$, and due to the high spatial resolution of
the STS technique, we were able to relate the local values of
$T_C$, $\Delta_\pi$ and $H_{c2}$. We have found a reduction of
$T_C$ for increasing doping, corresponding to a monotonous rising
of the BCS ratio but not of the absolute amplitude of the energy
gap. In agreement with recent theoretical models, we have measured
the largest gap value ($\Delta_\pi$ = 2.3 meV, 15\% larger than in
pure MgB$_2$) in samples with x=0.1, corresponding to a local
H$_{c2}^{\parallel c} \simeq$ 3T (30\% higher than in pure
MgB$_2$).


\acknowledgments
We thank  A. Bussmann-Holder and  A. Bianconi for
comments and useful discussions.


\thebibliography{apssamp}
\bibitem{Nagamatsu} J. Nagamatsu et al., \emph{Nature (London)}, \textbf{410}, 63 (2001).
\bibitem{Chol} H.J. Chol et al.,
\emph{Nature (London)}, \textbf{418}, 758 (2002).
\bibitem{An} J.M. An, W.E. Pickett, \emph{Phys. Rev. Lett.}, \textbf{86},
  4366 (2001).
\bibitem{Kortus} J. Kortus et al., \emph{Phys. Rev. Lett.} \textbf{86}, 4656
(2001).
  \bibitem{Budko} S.L. Budko $\it{et\,al.}$, \emph{Phys. Rev. Lett.} \textbf{86}, 1877 (2001)
\bibitem{Liu} A.Y. Liu et al., \emph{Phys. Rev. Lett.} \textbf{87},
087005 (2001).
\bibitem{Giubileo1} F. Giubileo et al., \emph{Europhys. Lett.} \textbf{58},
764 (2002).
\bibitem{GiubileoPRL} F. Giubileo et el., \emph{Phys. Rev. Lett.}, \textbf{87},
  177008 (2001).
\bibitem{Szabo} P. Szabo et al., \emph{Phys. Rev. Lett.} \textbf{87},  137005 (2001).
\bibitem{Gonnelli1} R.S. Gonnelli et al., \emph{Phys. Rev. Lett.}
\textbf{89}, 247004 (2002).
\bibitem{Iavarone1} M. Iavarone et al., \emph{Phys. Rev. Lett.} \textbf{89},
187002 (2002).
\bibitem{PhysC} For a review, see: \emph{Physica C: Superconductivity}, Volume 385, Issues
1-2(2003), edited by G. Crabtree, W. Kwok, S.L. Bud'ko and P.C.
Canfield.
\bibitem{sluski} J.S. Slusky et al.,  \emph{Nature (London)}, \textbf{410}, 343 (2001).
\bibitem{ribeiro}
R. A. Ribeiro et al., \emph{Physica C}, \textbf{384},
  227 (2003).
\bibitem{Erwin} S. C. Erwin, I. I. Mazin,
\emph{Phys. Rev. B}, \textbf{68}, 132505 (2003).
\bibitem{BH} A. Bussmann-Holder, A. Bianconi, \emph{Phys. Rev. B}, \textbf{67},
132509 (2003).
\bibitem{bianconi}
A. Bianconi et al., \emph{Phys. Rev. B}, \textbf{65}, 174515
(2002).
\bibitem{postorino} P. Postorino et al., \emph{Phys. Rev. B},
\textbf{65}, 020507 (2002).
\bibitem{renker} B. Renker et al., \emph{Phys. Rev.
Lett}, \textbf{88}, 067001 (2002).
\bibitem{PuttiAff}
M. Putti et al., \emph{Phys. Rev. B}, \textbf{68}, 094514 (2003).
\bibitem{PuttiGon} M. Putti et al., \emph{Phys. Rev. B}, \textbf{71}, 144505 (2005).
\bibitem{Wang} Y. Wang et al., \emph{J. Phy. Condens. Matter}, \textbf{15}, 883 (2003).
\bibitem{Putti06}
M. Putti et al., \emph{Phys. Rev. Lett}, \textbf{96}, 077003
(2006).
\bibitem{Iavarone} M. Iavarone et al., \emph{Phys. Rev. B}, \textbf{71}, 214502 (2005).
\bibitem{Gonnelli}
D. Daghero et al., \emph{Phys. Stat. Sol.}, \textbf{2},
  1656 (2005).
\bibitem{Karpinski}
J. Karpinski et al., \emph{Phys. Rev. B}, \textbf{71},
  174506 (2005).
  \bibitem{Kortus2} J. Kortus et al., Phys. Rev. Lett. \textbf{94}, 027002 (2005).
\bibitem{reply1} P. Samuely et al., Phys. Rev. Lett. \textbf{94}, 099701 (2005).
\bibitem{reply2} J. Kortus et al. Reply, Phys. Rev. Lett. \textbf{94}, 099702 (2005).
\bibitem{13}J. Karpinski et al., \emph{Supercond. Sci. Technol.},
\textbf{16}, 221 (2003); J. Karpinski et al., \emph{Physica C},
\textbf{385}, 42 (2003).
\bibitem{Dynes}
R.C. Dynes et al., \emph{Phys. Rev. Lett.}, \textbf{41}, 1509
(1978).
\bibitem{APL} A. Kohen et al., Appl. Phys. Lett. \textbf{86}, 212503
(2005).
\bibitem{Qubit} R. Cubitt et al., Phys. Rev. Lett. \textbf{91}, 047002 (2003).
\bibitem{Suhl} H. Suhl et al., Phys. Rev. Lett.
\textbf{3}, 552 (1959).
\end{document}